%% file: main.tex
\documentclass[sigconf,natbib=true]{acmart}

\usepackage{textcomp}
\usepackage{stfloats}
\usepackage{url}
\usepackage{verbatim}
\usepackage{amsthm}
\usepackage{subfigure}
\usepackage{makecell}
\usepackage{multirow}
\usepackage{enumitem}
\usepackage{subcaption}

\usepackage{color, colortbl}

\definecolor{Gray}{rgb}{0.9, 0.9, 0.9}

\newcommand{\gray}{\cellcolor{Gray}}

\usepackage[capitalize,noabbrev]{cleveref}

\usepackage{times}
\usepackage[ruled,linesnumbered]{algorithm2e}

\usepackage[inkscapelatex=false]{svg}

\copyrightyear{2026}
\acmYear{2026}
\setcopyright{cc}
\setcctype{by}
\acmConference[SIGIR '26]{Proceedings of the 49th International ACM
SIGIR Conference on Research and Development in Information
Retrieval}{July 20--24, 2026}{Melbourne, VIC, Australia}
\acmBooktitle{Proceedings of the 49th International ACM SIGIR Conference
on Research and Development in Information Retrieval (SIGIR '26), July
20--24, 2026, Melbourne, VIC, Australia}
\acmDOI{10.1145/3805712.3809623}
\acmISBN{979-8-4007-2599-9/2026/07}

\begin{document}

\title[SG-URInit]{Well Begun is Half Done: Training-Free and Model-Agnostic Semantically Guaranteed User Representation Initialization for Multimodal Recommendation}

\author{Jinfeng Xu}
\email{jinfeng@connect.hku.hk}
\affiliation{%
  \institution{The University of Hong Kong}
  \city{HongKong SAR}
  \country{China}}

\author{Zheyu Chen}
\email{zheyu.chen@bit.edu.cn}
\affiliation{%
  \institution{Beijing Institute of Technology}
  \city{Beijing}
  \country{China}}

\author{Shuo Yang}
\email{shuoyang.ee@gmail.com}
\affiliation{%
  \institution{The University of Hong Kong}
  \city{HongKong SAR}
  \country{China}}

\author{Jinze Li}
\email{lijinze-hku@connect.hku.hk}
\affiliation{%
  \institution{The University of Hong Kong}
  \city{HongKong SAR}
  \country{China}}

\author{Hewei Wang}
\email{heweiw@andrew.cmu.edu}
\affiliation{%
    \institution{Carnegie Mellon University}  
    \city{Pittsburgh, PA}   
    \country{USA}}

\author{Jianheng Tang}
\email{tangentheng@gmail.com}
\affiliation{%
    \institution{Peking University}  
    \city{Beijing}   
    \country{China}}

\author{Wei Wang}
\email{weiwang@mpu.edu.mo}
\affiliation{%
  \institution{Macao Polytechnic University}
  \city{Macao SAR}
  \country{China}}

\author{Xiping Hu}
\email{huxp@bit.edu.cn}
\affiliation{%
  {\institution{Beijing Institute of Technology}}
  \city{Beijing}
  \country{China}}

\author{Edith C. H. Ngai}
\authornote{Corresponding authors}
\email{chngai@eee.hku.hk}
\affiliation{%
  \institution{The University of Hong Kong}
  \city{HongKong SAR}
  \country{China}}

\renewcommand{\shortauthors}{Jinfeng Xu et al.}

 
\begin{abstract}
Recent advancements in multimodal recommendations, which leverage diverse modality information to mitigate data sparsity and improve recommendation accuracy, have gained significant attention. However, existing multimodal recommendations overlook the critical role of user representation initialization. Unlike items, which are naturally associated with rich modality information, users lack such inherent information. Consequently, item representations initialized based on meaningful modality information and user representations initialized randomly exhibit a significant semantic gap. 

To this end, we propose a Semantically Guaranteed User Representation Initialization (SG-URInit). SG-URInit constructs the initial representation for each user by integrating both the modality features of the items they have interacted with and the global features of their corresponding clusters. SG-URInit enables the initialization of semantically enriched user representations that effectively capture both local (item-level) and global (cluster-level) semantics.

Our SG-URInit is training-free and model-agnostic, meaning it can be seamlessly integrated into existing multimodal recommendation models without incurring any additional computational overhead during training. Extensive experiments on multiple real-world datasets demonstrate that incorporating SG-URInit into advanced multimodal recommendation models significantly enhances recommendation performance. Furthermore, the results show that SG-URInit can further alleviate the item cold-start problem and also accelerate model convergence, making it an efficient and practical solution for multimodal recommendations.
\end{abstract}



\maketitle

\input{Tex/1.Introduction}

\input{Tex/2.Investigation}
\input{Tex/3.Methodology}
\input{Tex/4.Evaluation}
\input{Tex/5.Related_Work}
\input{Tex/6.Conclusion}

\begin{acks}
This work was supported by the UGC General Research Fund no. 17209822 and the Innovation and Technology Commission Fund no. ITS/383/23FP from Hong Kong.
\end{acks}

\bibliographystyle{ACM-Reference-Format}
\bibliography{reference}

\appendix

\input{Tex/7.Appendix}
\end{document}

%% file: Tex/1.Introduction.tex
\section{Introduction}
The rapid growth of the internet has resulted in an overwhelming abundance of information, making recommender systems crucial for filtering and personalizing content. Traditional systems rely on historical user-item interactions \cite{he2020lightgcn,xu2024fourierkan,xu2024aligngroup,chen2025squeeze}, but their performance is often limited by data sparsity, as they struggle to effectively model user preferences and item properties from limited interaction data. To address this, recent works have incorporated multimodal information \cite{xu2025mdvt,xu2025cohesion}, such as images and textual descriptions, which explicitly enrich user and item representations and alleviate data sparsity. Early work, such as VBPR \cite{he2016vbpr}, incorporated visual content with matrix factorization to improve recommendations, while later studies \cite{chen2019personalized,liu2019user,yu2023multi,xu2025mentor} combined visual and textual modalities to enhance item representations. Graph-based methods, such as MMGCN \cite{wei2019mmgcn}, introduced GCNs to model user-item interactions, and GRCN \cite{wei2020graph} further refined edge pruning to reduce noise. Additional advancements, such as DualGNN \cite{wang2021dualgnn} and LATTICE \cite{zhang2021mining}, leveraged item-item and user-user homogeneous graphs, while FREEDOM \cite{zhou2023tale} stabilized item representations. Recent studies, including MMSSL \cite{wei2023multi} and MICRO \cite{zhang2022latent}, adopted contrastive learning to enhance modality-specific representation, while BM3 \cite{zhou2023bootstrap} and MENTOR \cite{xu2025mentor} further align representations among different modalities. Furthermore, LGMRec \cite{guo2024lgmrec} and DiffMM \cite{jiang2024diffmm} utilized hypergraphs and diffusion models, and COHESION \cite{xu2025cohesion} introduced dual-stage modality fusion for enhanced representations.

Despite these advancements, a critical limitation remains largely unaddressed in the field of multimodal recommendation: the initialization of user representations. Unlike items, which are enriched with modality information, user representations are typically initialized randomly, creating a significant semantic gap between user and item embeddings. This gap introduces noise during graph convolution, hindering the learning of effective representations. Since GCNs update node representations by aggregating their neighbors' representations, the process of aggregating pre-trained, semantically guaranteed item embeddings with randomly initialized user embeddings becomes problematic. In this case, the random user embeddings act as sources of noise, propagating into the semantically guaranteed item representations during the initial training phases. In Section~\ref{sec:investigation}, we provide a detailed investigation to validate these observations.

\begin{figure*}[!t]
\centering
    \subfigure[MENTOR] {
        \label{fig:investigation1}
        \includegraphics[width=0.48\linewidth]{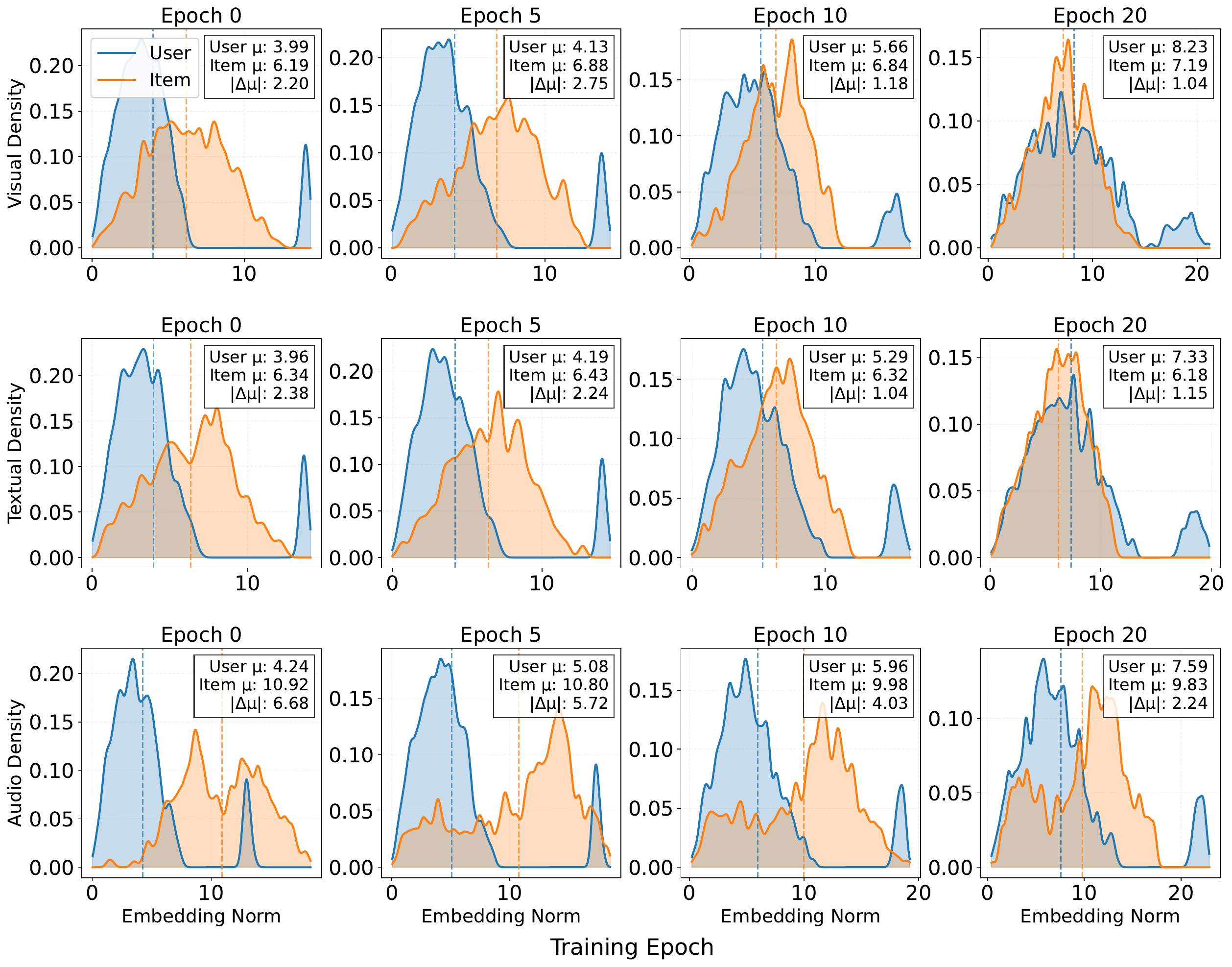}
        } 
    \subfigure[MENTOR with Initialization] {
        \label{fig:investigation2}
        \includegraphics[width=0.48\linewidth]{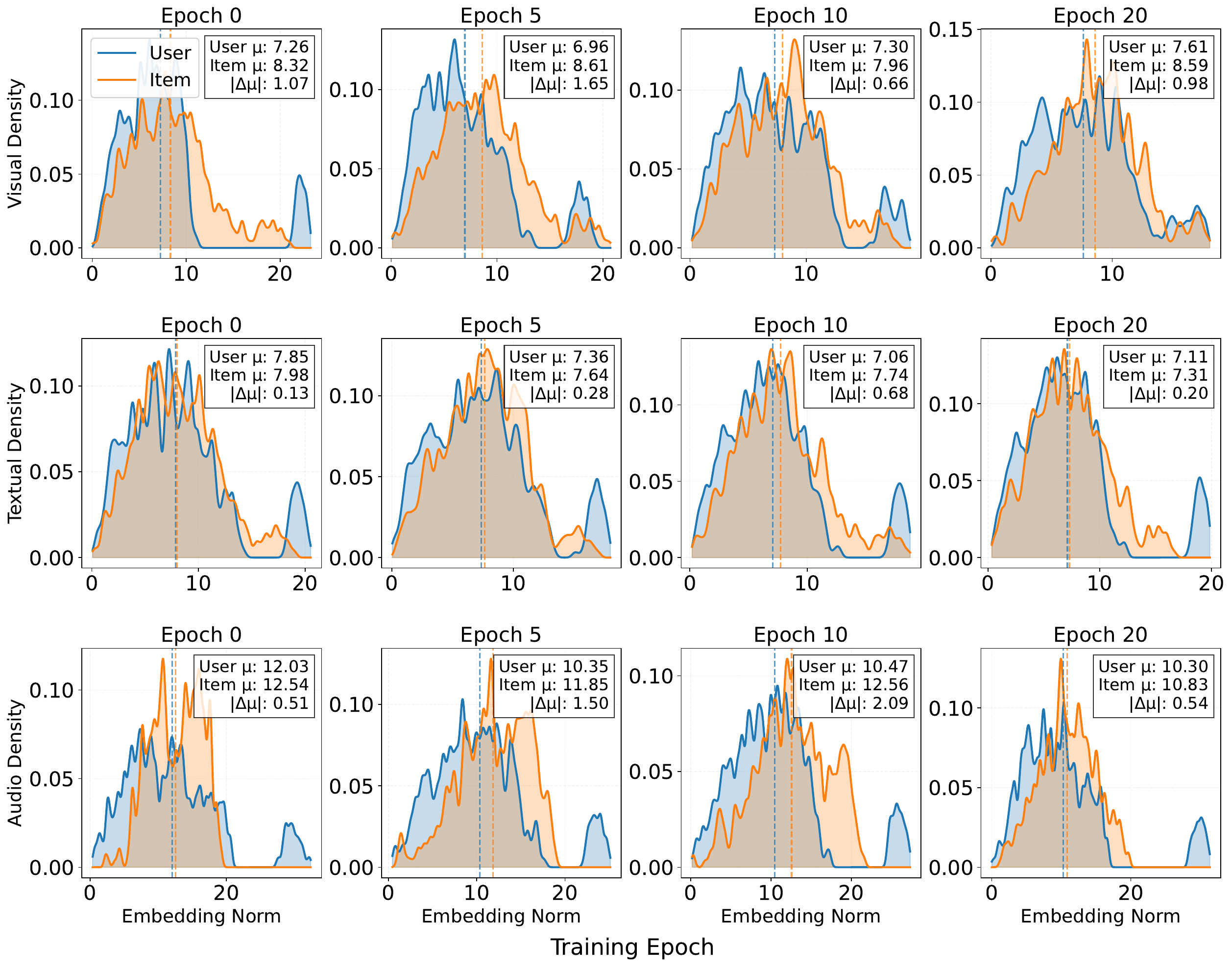}
        }  
    \vskip -0.2in
    \caption{Evolution of discrepancies between user and item representation during model optimization.}   
    \label{fig:investigation}   
\end{figure*}

While a naive solution might be to simply average the embeddings of interacted items, this solution suffers from local overfitting, restricting the user's representation to only their historical observations. To this end, we propose Semantically Guaranteed User Representation Initialization (SG-URInit). SG-URInit initializes semantically guaranteed user representations by integrating modality information of interacted items and cluster-level global information, effectively capturing both local (item-level) and global (cluster-level) semantics. This combination allows us to initialize a user representation that not only captures specific, localized preferences but also broadens the scope to include generalized preferences toward similar items within clusters. Our SG-URInit is training-free and model-agnostic, enabling seamless integration into existing multimodal recommendation models with no additional computational overhead. Extensive experiments on real-world datasets across various advanced multimodal recommendations demonstrate that SG-URInit significantly enhances recommendation performance, alleviates the item cold-start problem, and accelerates model convergence, establishing it as an efficient and practical solution for multimodal recommendations. In summary, our contributions can be outlined as follows:
\begin{itemize}[leftmargin=*]
    \item This work first reveals that the critical limitation remains largely unaddressed in multimodal recommendations: the initialization of user representations
    \item This work proposes SG-URInit, a semantically guaranteed user representation initialization, which mitigates the semantic gap between item and user initial representations by initializing a user representation that not only captures specific, localized preferences but also broadens the scope to include generalized preferences toward similar items within clusters.
    \item SG-URInit is training-free and model-agnostic, allowing seamless integration into existing multimodal recommendation models without additional computational overhead.
    \item We conduct comprehensive experiments on real-world datasets across various advanced multimodal recommendations, demonstrating that SG-URInit significantly enhances recommendation performance, alleviates the item cold-start problem, and accelerates model convergence.
\end{itemize}

%% file: Tex/2.Investigation.tex
\section{Investigation}
\label{sec:investigation}
In this section, we first provide a brief overview of the standard multimodal recommendation process. Then, we conducted an investigation into the discrepancies between user and item representations across different modalities during the model optimization process under two initialization settings: (1) random initialization of user representations and (2) initialization of user representations based on the representations of interacted items and clusters.

\subsection{Preliminary}
Given a user set $\mathcal{U}=\{u_1,...,u_{|\mathcal{U}|}\}$ and an item set $\mathcal{I}=\{i_1,...,i_{|\mathcal{I}|}\}$. In multimodal scenarios, each item contains multiple features, we introduce modality-specific item embedding $i^m$ for each item $i$ belonging to the set of modalities $\mathcal{M}$. The user-item interaction matrix is denoted as $\mathcal{R} \in \{0,1\}^{|\mathcal{U}| \times |\mathcal{I}|}$. Specifically, each entry $\mathcal{R}_{u,i}$ indicates whether the user $u$ is connected to item $i$, with a value of 1 representing a connection and 0 otherwise. These user-item interactions  $\mathcal{R}$. naturally constructs the bipartite graph $\mathcal{G} = (\mathcal{U}, \mathcal{I}, \mathcal{E})$, where $\mathcal{U}, \mathcal{I}$ serve as the graph vertices, and $\mathcal{E}$ denotes the edge set. For each user-item pair $(u,i)$ that satisfies $\mathcal{R}_{u,i} = 1$, there exists bidirectional edges $(u,i) \in \mathcal{E}$ and $(i,u) \in \mathcal{E}$. Previous works (refer to survey \cite{xu2025survey}) commonly random initialize $\mathbf{E}_{u}^{m} = \{\mathbf{e}_{u_1}^{m},\ldots,\mathbf{e}_{u_{|\mathcal{U}|}}^{m}\} \in \mathbb{R}^{d^m \times |\mathcal{U}|}$ to represent user representation with modality $m$. $\mathbf{E}_{i}^{m} = \{\mathbf{e}_{i_{1}}^{m},\ldots,\mathbf{e}_{i_{|\mathcal{I}|}}^{m}\}\in \mathbb{R}^{d^m \times|\mathcal{I}|}$ represents item initialized representation with modality $m$, which extracted by pre-trained encoders. Here $d^m$ represents the hidden dimensionality.

Subsequently, the initialized user and item representations are enhanced through various structures, such as homogeneous graphs \cite{he2020lightgcn,xu2024fourierkan,wei2019mmgcn} naturally formed by user-item interactions, heterogeneous graphs \cite{xu2025mentor,xu2025cohesion,zhou2023tale,zhang2021mining} constructed based on user preferences and item properties, or other advanced structures \cite{guo2024lgmrec} and techniques \cite{jiang2024diffmm}. These representations are then optimized using a combination of supervised and self-supervised objective functions. However, the semantic gap between randomly initialized user representations and modality-informed item representations adversely impacts both the optimization process and the final representation quality. We provide empirical support in subsequent subsection.

\subsection{Validation}
We selected the TikTok dataset, which contains visual, textual, and audio modalities (refer to Table~\ref{tab:dataset_statistics}), and used MENTOR, a state-of-the-art multimodal recommendation model, for our experiments. The experiments involved two variants: MENTOR and MENTOR with Initialization. The former adopts random initialization for user representations, as in existing multimodal recommendation models, while the latter initializes user representations based on the representations of interacted items (refer to Eq.~\ref{eq:item-level}).

We apply t-SNE \cite{van2008visualizing} to map the user and item representations into  two-dimensional space for each modality and plotted the distribution of their representation magnitudes (L2 norms). Figure~\ref{fig:investigation} illustrates the distributions of user and item representations across the three modalities at 0, 5, 10, and 20 epochs. By comparing the two curves (blue for users and orange for items), the similarities and discrepancies in their magnitude distributions can be observed. Additionally, each subfigure provides: (1) User $\mu$: the mean magnitude of user representations. (2) Item $\mu$: the mean magnitude of item representations. (3) $|\Delta \mu|$: the absolute difference between the two means. Dashed lines indicate the positions of the mean values (blue dashed line for users and orange dashed line for items). From the results, we observe that user representations initialized based on item representations are more similar to item representations than randomly initialized user representations, regardless of the training stage. It is worth noting that these representations are enhanced through techniques such as heterogeneous and homogeneous graph convolutions. This indicates that the semantic gap caused by initialization persists even after multiple layers of convolution, ultimately impacting the overall performance of the model.

To this end, we propose a tailored Semantically Guaranteed User Representation Initialization (SG-URInit) to bridge the semantic gap between user and item initial representations during initialization, ensuring a better optimization process and enhanced representations.

%% file: Tex/3.Methodology.tex
\begin{figure*}
    \centering
    \vskip -0.05in
    \includegraphics[width=1\linewidth]{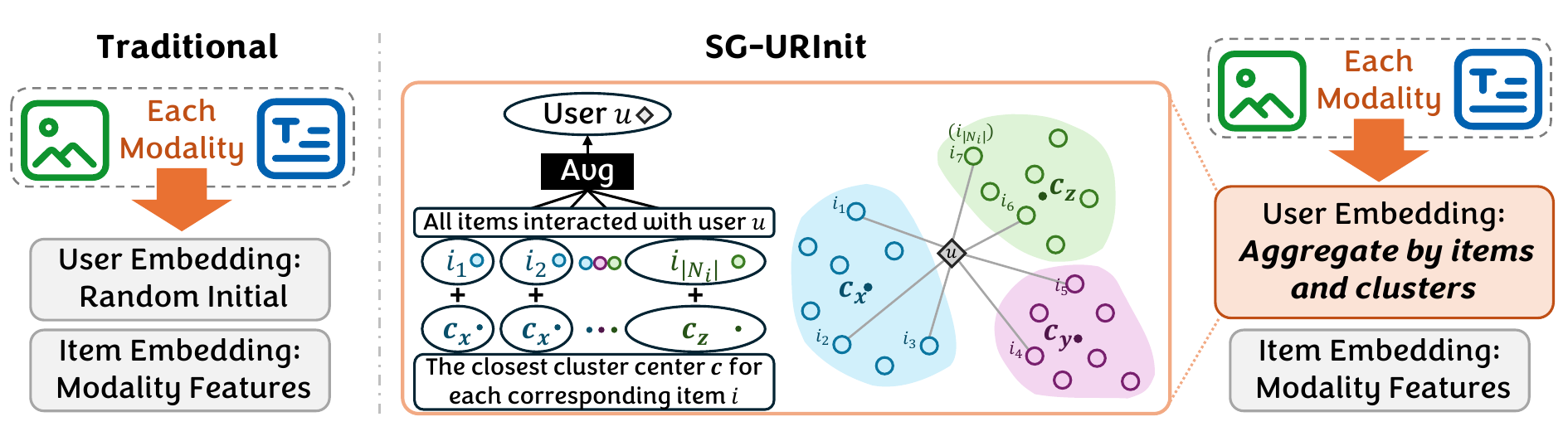}
    \vskip -0.15in
    \caption{Overview of user representation initialization. Left: traditional user representation initialization; Right: our SG-URInit.}
    \label{fig:SG-URInit}
\end{figure*}

\section{Methodology}
In Section~\ref{sec:investigation}, we discussed the drawbacks of the random user representation initialization commonly adopted in prior multimodal recommendation studies. While items are associated with specific modality information, users inherently lack such modality information. Consequently, item initialization leverages modality features, whereas user initialization typically relies on random values, resulting in a significant semantic gap. In this section, we detail our proposed Semantically Guaranteed User Representation Initialization (SG-URInit). SG-URInit constructs the initial representation for each user by integrating the modality information of the items the user has interacted with and the overall features of their corresponding clusters. SG-URInit enables the initialization of a semantically guaranteed user representation that effectively captures both local (item-level) and global (cluster-level) semantics. The overview of SG-URInit\footnote{Code is available as \href{https://github.com/Jinfeng-Xu/SG-URInit}{https://github.com/Jinfeng-Xu/SG-URInit}.} can be found in Figure~\ref{fig:SG-URInit}.

\subsection{SG-URInit}
The modality representation of a user should comprehensively reflect their modality-specific preference tendencies. A natural and intuitive approach is to aggregate the modality representations of items they have interacted with, which captures their local preferences at a item-level. However, relying solely on such aggregation risks overfitting to observed preferences, thereby neglecting the user's potential interest in similar but unobserved items. To address this limitation, SG-URInit further adopts a cluster-level aggregation. Firstly, we group items into clusters based on their modality features. Beyond aggregating the local modality representations of interacted items, we further incorporate the global cluster centroids of the corresponding items. This combination allows us to construct a user representation that not only captures specific, localized preferences but also broadens the scope to include generalized preferences toward similar items within clusters. By integrating both local and global semantic information, our SG-URInit ensures a more holistic and semantically guaranteed user representation.

\subsubsection{Item-level Aggregation}
We first perform item-level aggregation of the modality features of the items each user has interacted with. Formally:
\begin{equation}
\label{eq:item-level}
\bar{\mathbf{e}}_{u}^{m}=\sum_{\tilde{i} \in N(u)} \frac{1}{|N(\tilde{i})|} \mathbf{e}_{\tilde{i}}^{m},
\end{equation}
where $N(u)$ and $N(\tilde{i})$ refers to the set of items or users that interact with user $u$ and item $\tilde{i}$, respectively. The weight of each interacted item is influenced by its degree, inspired by the design of LightGCN \cite{he2020lightgcn}. We empirically validate in Section~\ref{sec:indepth} that this weight allocation strategy outperforms uniform weighting. 

\subsubsection{Cluster-level Aggregation}
Focusing solely on item-level aggregation risks overfitting to the user's observed preferences, which may overlook their potential interest in similar but unobserved items. To this end, SG-URInit introduces a cluster-level aggregation mechanism. We first partition all items into $K$ clusters based on their modality features through K-Means clustering:
\begin{equation}
\label{eq:KMeans}
\mathcal{C}^m=\operatorname{K-Means}(\{\mathbf{e}_i^m \mid \forall i \in \mathcal{I}\}, K),
\end{equation}
where $\mathcal{C}^m=\{c_1^m, c_2^m, \ldots, c_K^m\}$ denotes the cluster centroids for modality $m$. $K$ is the number of cluster centroids, which we empirically discussed in Section~\ref{sec:hyper}.

For each item $i$, we compute its cluster-aligned representation by assigning it to the nearest cluster centroid:
\begin{equation}
\tilde{\mathbf{e}}_i^m=\underset{c_k^m \in \mathcal{C}^m}{\arg \min }\left\|\mathbf{e}_i^m-c_k^m\right\|_2.
\end{equation}

The cluster-level user representation is then obtained by aggregating these cluster-aligned item representations:
\begin{equation}
\tilde{\boldsymbol{e}}_u^m=\sum_{\tilde{i} \in N(u)} \frac{1}{|N(\tilde{i})|} \tilde{\mathbf{e}}_{\tilde{i}}^m,
\end{equation}
where $N(u)$ and $N(\tilde{i})$ denote neighbor sets for user $u$ and item $\tilde{i}$, respectively. To ensure consistency, we adopt the same weighting strategy as that used at the item-level.

\subsection{Fusion}
We combine the user representations generated by item-level aggregation and cluster-level aggregation to form the final user representation, which simultaneously incorporates both local and global semantics. Formally:
\begin{equation}
\label{eq:fusion}
\mathbf{e}_{u}^{m} = (1-\lambda) \bar{\mathbf{e}}_{u}^{m} + \lambda \tilde{\boldsymbol{e}}_u^m,
\end{equation}
where $\lambda \in[0,1]$ is a hyper-parameter that adjusts the scaling and balances the contributions of the item-level ($\bar{\mathbf{e}}_u^m$) and cluster-level ($\tilde{\boldsymbol{e}}_u^m$) user representations. Since both $\bar{\mathbf{e}}_u^m$ and $\tilde{\boldsymbol{e}}_u^m$ share the same representation scalar level as the item representations, directly summing the two would lead to an increase in the scalar magnitude of the user representation. To this end, we introduce a scaling factor 
$\lambda$ to balance and normalize the aggregation, ensuring the final user representation maintains the same scalar level as the original representations.

\subsection{Discussion}
This form of initialization is conceptually analogous to the behavior of graph convolution, as the essence of representation enhancement in graph convolution lies in utilizing the similar representations of neighboring nodes (interacted nodes) to enrich a node's own representation. We point out that the effectiveness of graph convolution can be significantly improved by our semantically guaranteed user representation initialization.

Specifically, prior works often initialize item representations using meaningful modality information, while user representations are randomly initialized \cite{xu2025survey}. As a result, in a direct single-layer graph convolution operation, the aggregation of randomly initialized user representations for items can essentially be viewed as injecting random noise (randomly initialized user representation) into the item representations. On the other hand, the aggregation of modality-informed item representations for users can be interpreted as a process akin to our semantically guaranteed user representation initialization, but with the additional introduction of noise through the randomly initialized user representations. 

To address these shortcomings, our semantically guaranteed user representation initialization can effectively mitigate the propagation of noise into item representations during the convolution process, while also avoiding the influence of users' initial random noise. This ensures a more robust and semantically aligned representation learning process, enhancing both the quality of the representations and the overall performance of graph-based models.

%% file: Tex/4.Evaluation.tex
\section{Evaluation}
\label{sec:eval}
We conduct extensive experiments\footnote{To avoid the instability of K-Means, all results reported in this paper are averaged over 10 K-Means runs.} aiming to answer the following research questions: \textbf{RQ1:} Can SG-URInit enhance the performance of multimodal recommender systems? \textbf{RQ2:} How do the different components in SG-URInit affect performance enhancement? \textbf{RQ3:} Can SG-URInit have a positive impact on the convergence speed? \textbf{RQ4:} Can SG-URInit be compatible with robust training and data augmentation strategies? \textbf{RQ5:} Can SG-URInit enhance multimodal recommendations in different sparse data scenarios? \textbf{RQ6:} Does SG-URInit perform satisfactorily in the item cold-start scenario? \textbf{RQ7:} What is the impact of key hyper-parameters in SG-URInit? \textbf{RQ8:} What is the cost of user initialization in SG-URInit?

\input{Tab/Evaluation}
\input{Tab/Dataset}

\subsection{Settings}
\subsubsection{Datasets}
The experiments are conducted on three real-world datasets: Baby, Sports, and Clothing from Amazon \cite{mcauley2015image}. All the datasets comprise textual and visual features in the form of item descriptions and images. To further evaluate the performance of SG-URInit in scenarios involving multiple modalities, we also conduct experiments on the TikTok dataset \cite{jiang2024diffmm}. Our data preprocessing methodology follows the approach outlined in MMRec \cite{zhou2023mmrecsm}. Table~\ref{tab:dataset_statistics} shows the statistics of these datasets. 

\subsubsection{Baselines}
We extensively examine the performance of our SG-URInit across a variety of multimodal recommendation models: 
\begin{itemize}[leftmargin=*]
\item \textbf{MMGCN} \cite{wei2019mmgcn} applies GCN for each data modality to learn modality-specific features and then integrates all user-predicted ratings across modalities to produce the final rating.
\item \textbf{SLMRec} \cite{tao2022self} leverages a self-supervised learning framework for multimodal recommendations by establishing a tailored node self-discrimination task, which reveals hidden multimodal patterns.
\item \textbf{FREEDOM} \cite{zhou2023tale} refines LATTICE by freezing the item-item graph to stabilize item relationships and reducing noise in the user-item graph to enhance recommendation accuracy.
\item \textbf{DRAGON} \cite{zhou2023enhancing} leverages heterogeneous and homogeneous graphs to learn high-quality user/item representations.
\item \textbf{LGMRec} \cite{guo2024lgmrec} integrates local embeddings, which capture fine-grained topological embeddings, with global embeddings considering hypergraph dependencies among items.
\item \textbf{MENTOR} \cite{xu2025mentor} proposes multi-level cross-modal alignment tasks to effectively improve final representation and achieve state-of-the-art recommendation accuracy.
\end{itemize}
Moreover, we test the compatibility of our SG-URInit with the adversarial training strategy (AMR \cite{tang2019adversarial}) and LLM-based data augmentation strategy (GPT-4o \cite{yang2023dawn}).

\subsubsection{Metrics}
For a fair comparison, we follow the settings of previous works \cite{xu2025survey,zhou2023bootstrap,zhou2023tale} to adopt two widely-used evaluation metrics for top-$K$ recommendation: Recall@K and NDCG@K. We report the average scores for all users in the test dataset under both K = 5 and K = 10, respectively.

\subsubsection{Implementation Details}
We retain the standard settings for all baselines and fix batch size $B$ to $2048$. We perform a complete grid search for all baselines to find the optimal setting based on their published papers. For SG-URInit, we apply a grid search on hyper-parameter $K$ in Eq~\ref{eq:KMeans} in $\{2, 4, 8, 16\}$, and balancing hyper-parameter $\lambda$ in Eq.~\ref{eq:fusion} in $\{0, 0.0001, 0.001, 0.01, 0.1, 1\}$. The common optimizer is Adam \cite{kingma2014adam} and all training and evaluation of all models are conducted on an RTX4090 GPU. For the GPT-4o data augmentation strategy, we utilize GPT-4o \cite{yang2023dawn} to augment the raw description via the items' image for all datasets to improve the correlation between textual and visual modalities. We designed the prompt as: \textbf{`[$V$] Here is the description of an item and the corresponding picture, please combine the picture to improve the description quality in one paragraph. The description is as follows: [$T$].'}, where [$V$] and [$T$] are the raw image and description for each item, respectively.

\subsection{Overall Performance (RQ1)}

We evaluate the effectiveness of our SG-URInit on various models across multiple real-world datasets in multimodal recommendation scenarios. From Table~\ref{tab:comparison_results}, we find the following observations:

\noindent \textbf{\underline{Observation 1}: SG-URInit can enhance the performance of various multimodal recommendation models across all datasets.} As shown in Table~\ref{tab:comparison_results}, we conducted extensive experiments using SG-URInit on six multimodal recommendation models across four distinct public datasets. The experimental results demonstrate that SG-URInit achieves significant improvements over all baselines across all evaluation metrics. In summary, these results validate that leveraging item multimodal information to initialize user representations can effectively enhance recommender performance by mitigating the semantic gap between user and item representations.

\noindent \textbf{\underline{Observation 2}: SG-URInit is more effective on datasets with a greater variety of modality information.} The performance improvement brought by SG-URInit is particularly significant on the TikTok dataset, which we attribute to that TikTok contains a richer variety of modality information. In contrast, the random initialization of user representations faces more pronounced challenges when handling datasets with more diverse modality information. SG-URInit addresses this concern effectively, providing a promising solution for real-world scenarios that incorporate richer multimodal information.

\subsection{Ablation Study (RQ2)}
To discern the impact of our SG-URInit's core components, we conducted an ablation study with various configurations:
\begin{itemize}[leftmargin=*]
    \item $w/o$-Cluster: It removes the cluster-level aggregation, resulting in an user representation initialization with only local semantic. This configuration is equivalent to a recent study (AlphaRec \cite{AlphaRec2025}).
    \item $w/o$-Item: It removes the item-level aggregation, resulting in an user representation initialization with only global semantic. 
\end{itemize}

\begin{figure*}[!t]
    \centering 
    \includegraphics[width=1\linewidth]{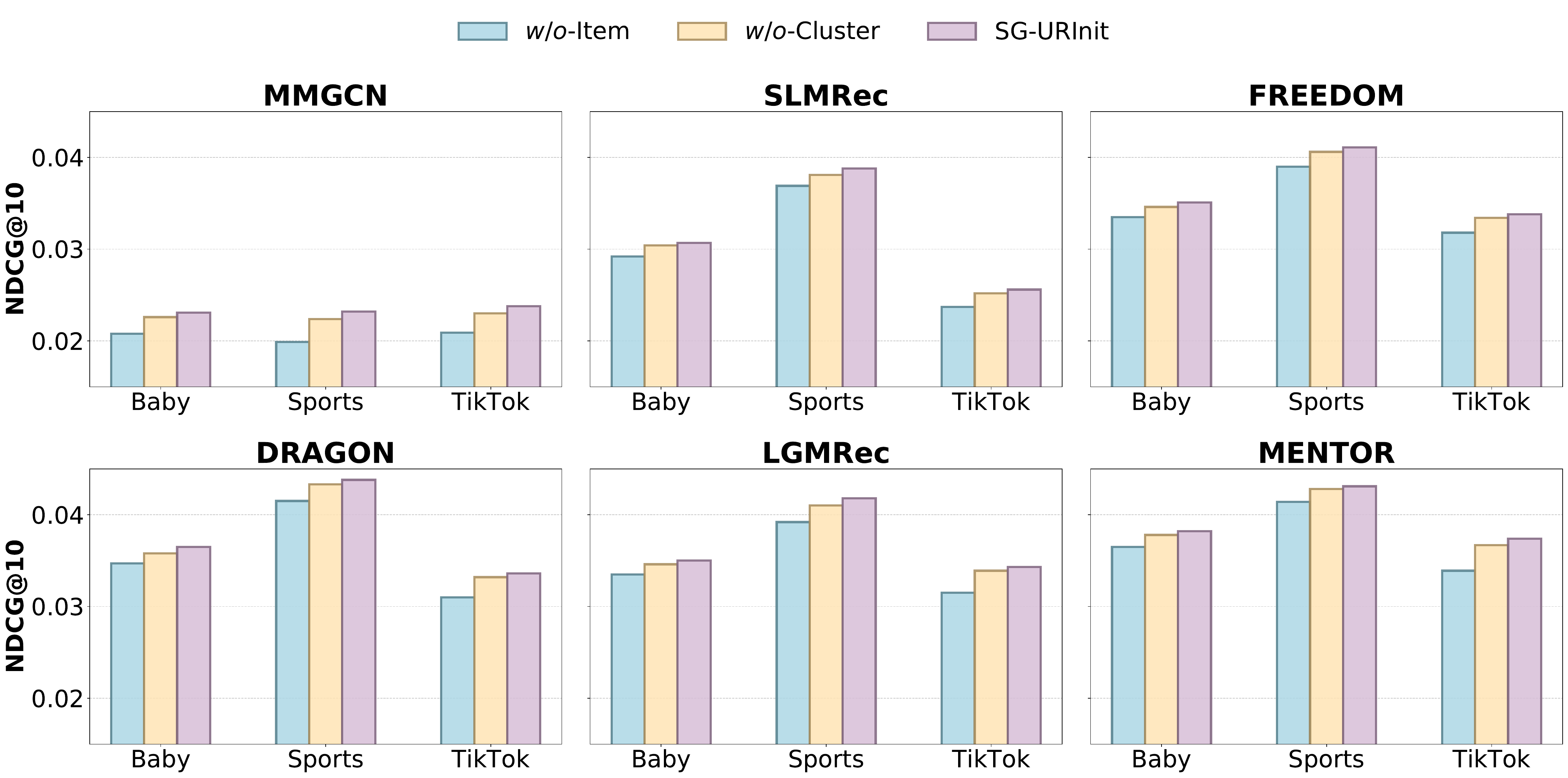}
     \vskip -0.1in
    \caption{Ablation study on key components of SG-URInit in terms of NDCG@10.}
    \label{fig:ab}
     \vskip -0.05in
\end{figure*}

We conduct extensive experiments for our SG-URInit across all six multimodal recommendation models on three datasets for various configurations. The findings presented in Figure~\ref{fig:ab} clearly demonstrate that our SG-URInit surpasses all its modified configurations, thereby confirming the essential role each component plays in initialing high-quality user representations. It is worth noting that $w/o$-Item consistently underperforms compared to $w/o$-Cluster, indicating that users' local multimodal information plays a dominant role. Meanwhile, the fact that $w/o$-Cluster consistently underperforms compared to SG-URInit demonstrates that global multimodal information also provides valuable semantic information for the initialization process.



\subsection{Convergence Speed (RQ3)}
\label{sec:convergence}
\begin{figure}[!t]
    \centering
    \includegraphics[width=1\linewidth]{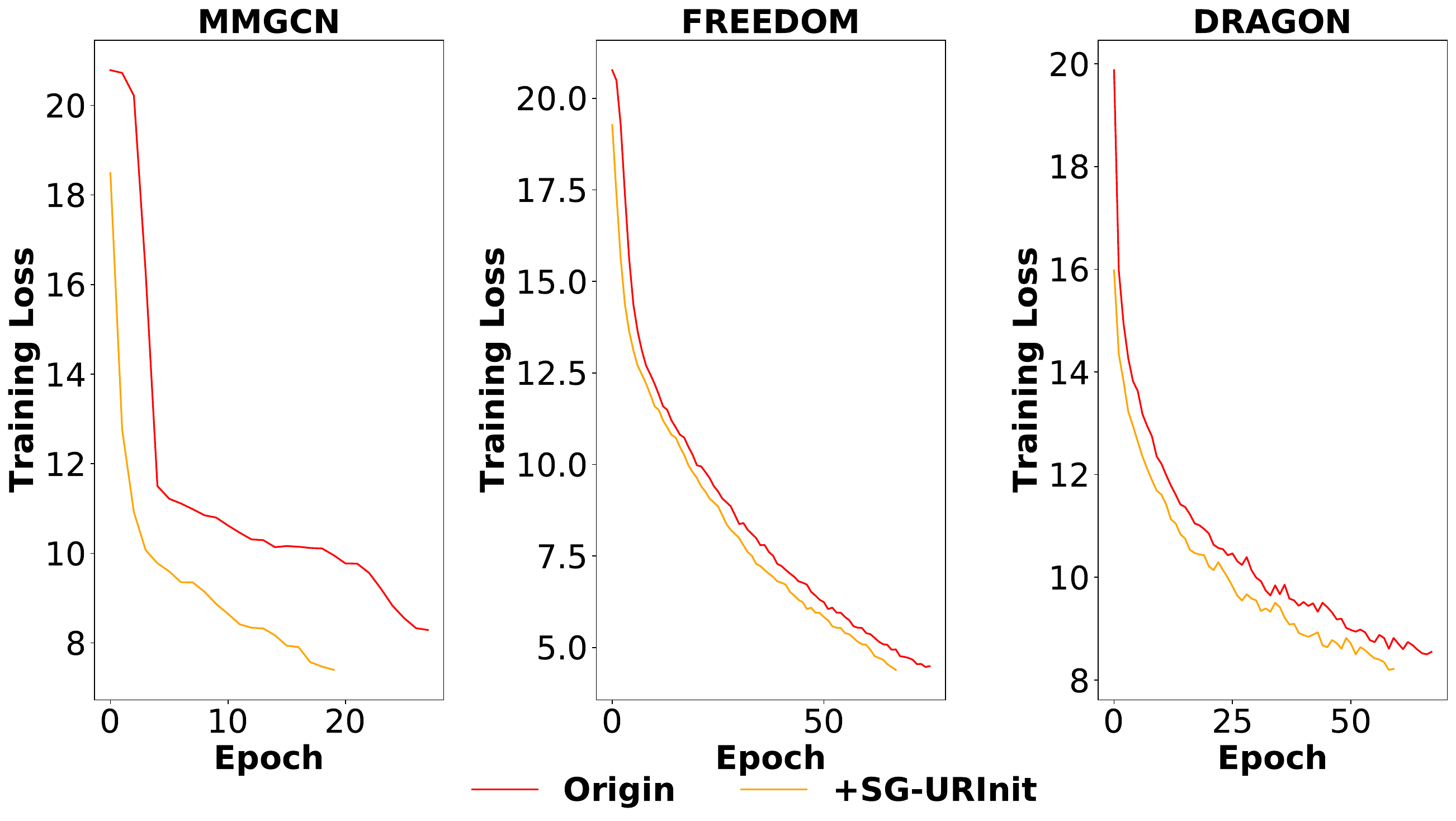}
    \vskip -0.1in
    \caption{Convergence study on the TikTok dataset.}
    \vskip -0.05in
    \vskip -0.1in
    \label{fig:convergence}
\end{figure}

In addition, SG-URInit helps accelerate model training convergence. We visualized the training loss of three advanced multimodal recommendation models (MMGCN, FREEDOM, and DRAGON) on the TikTok dataset. Following previous training settings \cite{zhou2023bootstrap,zhou2023tale}, we adopted an early stopping strategy with a patience of 20 epochs and set the maximum number of epochs to 1,000. As shown in Figure~\ref{fig:convergence}, SG-URInit effectively improves the convergence speed of all models. We attribute this improvement to the reduction in the gap between user and item representations achieved by our user representation initialization, which accelerates model training convergence.

\subsection{Compatibility Analysis (RQ4)}
\input{Tab/Compatibility}
Existing studies improve the robustness of multimodal recommendation systems through adversarial training strategies \cite{li2020adversarial, tang2019adversarial} and data augmentation methods \cite{luo2024integrating, huang2024large, wei2024llmrec}. Building on this, we further evaluate the compatibility of SG-URInit with the adversarial training strategy (AMR \cite{tang2019adversarial}) and the LLM-based data augmentation strategy (GPT-4o \cite{yang2023dawn}). We conduct extensive experiments on two multimodal recommendation models across three public datasets. As shown in Table~\ref{tab:compatibility}, combining SG-URInit with both AMR and GPT-4o further enhances model performance. Furthermore, GPT-4o consistently outperforms AMR across all datasets, which we attribute to its ability to better bridge the inherent gap among modalities information of items. Notably, using both AMR and GPT-4o simultaneously achieves the best performance, surpassing the results of employing either method individually.

\subsection{Sparsity Analysis (RQ5)}
\begin{figure*}[!t]
    \centering
    \includegraphics[width=1\linewidth]{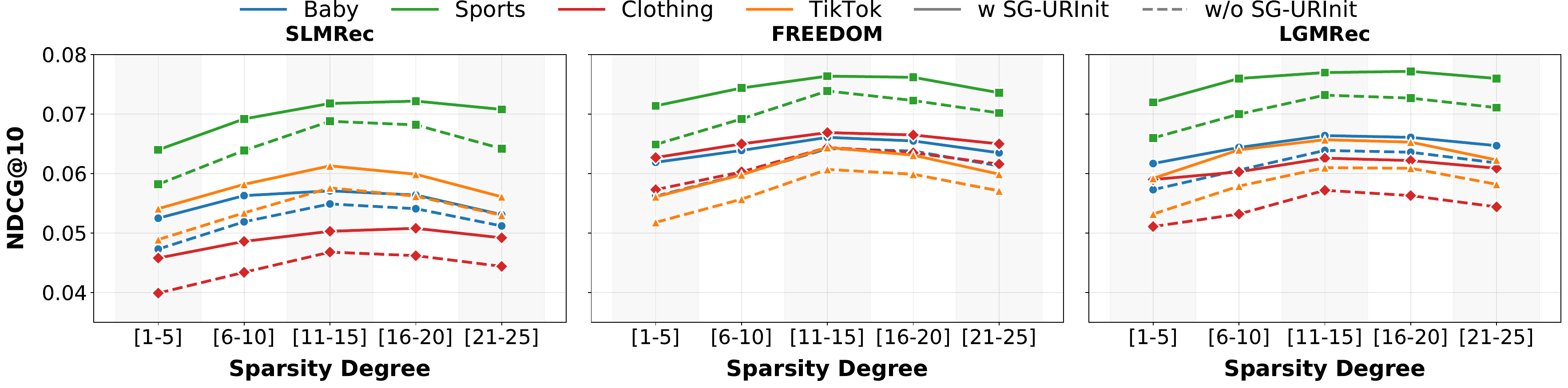}
     \vskip -0.1in
    \caption{Sparsity study on three advanced multimodal recommendation models across four distinct datasets.}
    \label{fig:sparsity}
\end{figure*}

To assess the impact of SG-URInit on advanced multimodal recommendation models under varying levels of data sparsity, we performed experiments on subsets of four datasets with different sparsity conditions. We compared the performance of three state-of-the-art multimodal recommendation models—SLMRec, FREEDOM, and LGMRec—both with and without the integration of SG-URInit. To better understand the effect of data sparsity, users were grouped based on the number of interactions in the training data (e.g., the first group includes users with 1–5 interactions). As shown in Figure~\ref{fig:sparsity}, SG-URInit consistently improves the performance of all models across datasets and user groups, confirming its robustness under sparse conditions. Notably, the performance boost provided by SG-URInit is especially prominent in highly sparse scenarios. This can be attributed to the fact that a well-initialized user representation reduces the dependency on extensive data during training in these challenging settings.

\subsection{Cold-Start Analysis (RQ6)}
\input{Tab/Cold_Start}
We report the results for the item cold-start scenario on the Baby and TikTok datasets. We adopt the widely used experimental settings described in \cite{zhang2022latent}. Specifically, we remove all user-item interaction pairs associated with a randomly selected 20\% item set from the training set. We further divide the half of the items (10\%) into the validation set and half (10\%) into the testing set. In other words, these items are entirely unseen in the training set. As shown in Table~\ref{tab:cold-start}, SG-URInit enhances the performance of multimodal recommendation models in item cold-start situations. This improvement can be attributed to the semantic consistency introduced by our user representation initialization, which narrows the semantic gap between user and item representations.

\subsection{Hyper-parameter Analysis (RQ7)}
\label{sec:hyper}
\begin{figure*}[!t]
    \centering
    \includegraphics[width=1\linewidth]{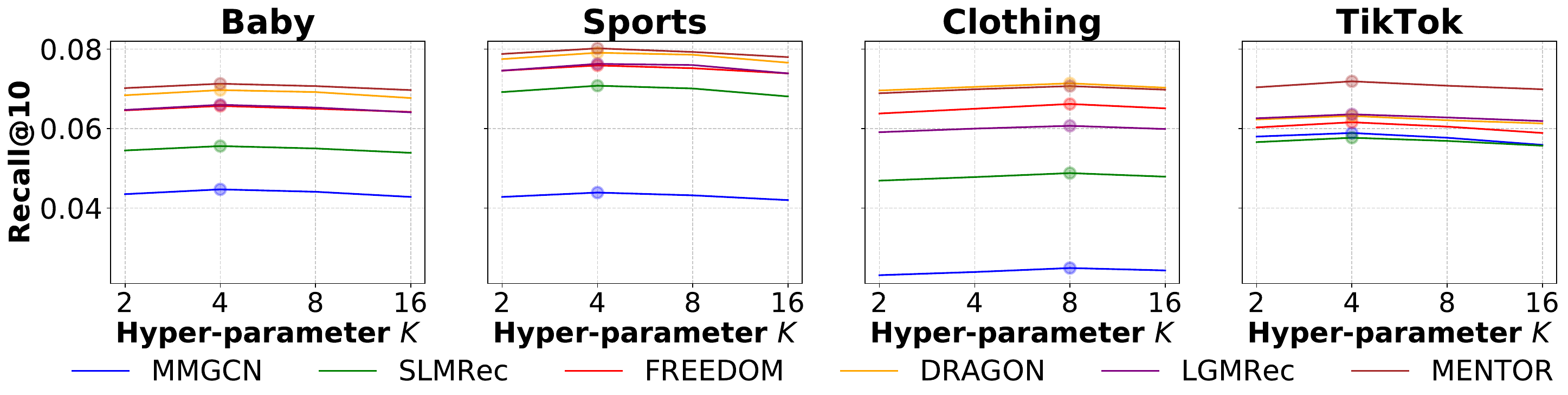}
    \vskip -0.1in
    \caption{Performance $w.r.t.$ hyper-parameter $K$.}
    \label{fig:Hyper1}
\end{figure*}

\begin{figure*}[!t]
    \centering
    \includegraphics[width=1\linewidth]{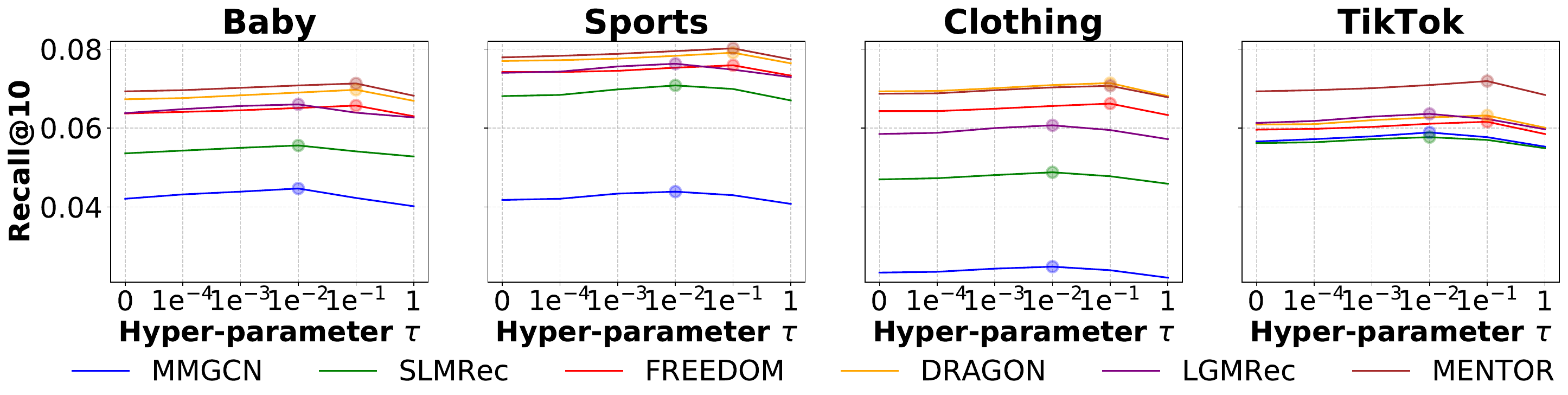}
    \vskip -0.1in
    \caption{Performance $w.r.t.$ hyper-parameter $\lambda$.}
    \label{fig:Hyper2}
\end{figure*}

We evaluate the impact of the key hyper-parameters ($K$ and $\lambda$) on MDVT's performance across four distinct datasets in terms of Recall@10. Results are presented in Figure~\ref{fig:Hyper1} and Figure~\ref{fig:Hyper2}.


\noindent\textbf{Hyper-parameter $K$}: In Figure~\ref{fig:Hyper1}, we observe that for hyper-parameter $K$, all models require the same optimal setting on each dataset, although the optimal setting varies across different datasets. The choice of $K$ depends on the complexity of the scenario: for the more complex Clothing dataset, $K = 8$ is the best choice, while for the other datasets, $K = 4$ yields the best performance.

\noindent\textbf{Hyper-parameter $\lambda$}: In Figure~\ref{fig:Hyper2}, we observe that for hyper-parameter $\lambda$, the optimal value is consistent across different models and less influenced by the dataset. For models incorporating the item-item graph (FREEDOM, DRAGON, and MENTOR), the optimal $\lambda$ is $0.1$, whereas for other models, the optimal value is $\lambda = 0.01$. Notably, when $\lambda = 0$, the SG-URInit degenerates into the variant w/o-Cluster, and when $\lambda = 1$, it degenerates into the variant w/o-Item.

\noindent \textit{Insights}: We further provide guidance on the hyper-parameter settings for SG-URInit. For the hyper-parameter $K$, setting $K = 4$ consistently yields satisfactory performance improvements across all models and datasets. In more complex scenarios, exploring $K = 8$ may offer additional performance gains. For the hyper-parameter $\lambda$, we recommend using $\lambda = 0.1$ for models incorporating item-item graphs, while $\lambda = 0.01$ is more suitable for other models. These insights significantly reduce the cost of hyper-parameter tuning when implementing SG-URInit.

\input{Tab/Appendix1}

\input{Tab/Cost}

\input{Tab/Appendix2}

\subsection{Discussion for Training-Free and Model-Agnostic}
SG-URInit features two key advantages: it is training-free and model-agnostic. Its initialization is pre-constructed based on the interaction data and multimodal information of items in each dataset, requiring no additional overhead during the training phase. The user representations initialized for each modality can be directly applied to any existing multimodal recommendation model. Notably, as demonstrated in Section~\ref{sec:convergence}, SG-URInit not only imposes no burden on the training process but also accelerates model convergence, thereby reducing the overall training time.

\subsection{Cost of User Initialization (RQ8)}
\label{sec:init}
We report the pre-training user initialization times for four datasets, specifically focusing on $K = 4$ and $K = 8$, as these settings consistently delivered the best performance. The results, summarized in Table~\ref{tab:cost}, show the time costs associated with user initialization. Notably, the user initialization process is conducted entirely before the training phase and can be reused across multiple runs, significantly reducing the overall impact of these time costs.

\subsection{In-depth Analysis}
\label{sec:indepth}
Ablation studies on weighting strategies during aggregation and the impact of setting different numbers of cluster centers $K$ for different modalities on SG-URInit, are topics worthy of further exploration.
\subsubsection{Ablation Study}
\label{appendix:ablation}
We conduct an empirical discussion on the weight allocation strategies for aggregating items or clusters during user representation initialization. Specifically, in addition to the degree-based aggregation used by SG-URInit, we also adopt other strategies, such as bi-degree weighting (as in LightGCN) and equal weighting. The detailed formulas are as follows: 
\begin{itemize}[leftmargin=*]
    \item SG-URInit: $\sum_{\tilde{i} \in N(u)} \frac{1}{|N(\tilde{i})|}$.
    \item -bi-degree: $\sum_{\tilde{i} \in N(u)} \frac{1}{\sqrt{|N(\tilde{i})||N(u)|}}$.
    \item -equal $\sum_{\tilde{i} \in N(u)} \frac{1}{|N(u)|}$.
\end{itemize}

Table~\ref{tab:ablation2} presents the performance comparison of SG-URInit and other weight allocation strategy variants across three datasets and three different advanced models. We observe that SG-URInit generally outperforms all variants, while the bi-degree weighting strategy also significantly outperforms the equal weighting strategy. We attribute this to the importance of popularity-based weighting in the recommendations.

\subsubsection{Hyper-parameter Analysis}
\label{appendix:hyper}

We further explore whether setting different numbers of cluster centers $K$ for different modalities can enhance the performance of SG-URInit. The value of $K$ for each modality is searched within $\{2, 4, 8, 16\}$. We report the optimal performance under dynamic $K$ for all four datasets across four models, along with the corresponding $K$ selection for each modality.

According to the experimental results in Table~\ref{tab:ablation3}, we find that dynamic control of $K$ achieves better performance for datasets other than Clothing. We attribute this to the varying levels of information richness across different modalities. For instance, the textual modality generally contains more fine-grained information than the visual modality, and the degree of granularity for each modality varies across different scenarios. Therefore, dynamically adjusting $K$ can further enhance the performance of SG-URInit.


%% file: Tab/Evaluation.tex
\begin{table*}[!t]
\caption{Performance comparison of baselines with or without SG-URInit on all datasets in terms of Recall@K (R@K) and NDCG@K (N@K). $^*$ indicates the improvement is statistically significant, where the p-value is less than 0.01.}
 \vskip -0.05in
\centering
\tabcolsep=0.04in
\label{tab:comparison_results}
\resizebox{\linewidth}{!}{
    \begin{tabular}{c|cccc|cccc|cccc|cccc}
     \toprule
         Datasets&  \multicolumn{4}{c|}{Baby}&  \multicolumn{4}{c|}{Sports}&  \multicolumn{4}{c|}{Clothing}&  \multicolumn{4}{c}{TikTok}\\\midrule
         Metrics& R@5& R@10& N@5& N@10 & R@5& R@10& N@5& N@10& R@5& R@10& N@5& N@10& R@5& R@10& N@5& N@10\\\midrule
         MMGCN & 0.0240& 0.0378& 0.0160& 0.0200& 0.0216& 0.0370& 0.0143& 0.0193& 0.0130& 0.0218& 0.0088& 0.0110& 0.0250& 0.0485& 0.0125& 0.0200\\
         +SG-URInit & \textbf{0.0263$^*$}& \textbf{0.0447$^*$}& \textbf{0.0171$^*$}& \textbf{0.0231$^*$}& \textbf{0.0268$^*$}& \textbf{0.0439$^*$}& \textbf{0.0175$^*$}& \textbf{0.0232$^*$}& \textbf{0.0147$^*$}& \textbf{0.0249$^*$}& \textbf{0.0100$^*$}& \textbf{0.0126$^*$}& \textbf{0.0341$^*$}& \textbf{0.0589$^*$}& \textbf{0.0162$^*$}& \textbf{0.0238$^*$}\\
         \gray Improv. &\gray 9.58\%&\gray 18.25\%&\gray 6.88\%&\gray 15.50\%&\gray 24.07\%&\gray 18.65\%&\gray 22.38\%&\gray 20.21\%&\gray 13.08\%&\gray 14.22\%&\gray 13.64\%&\gray 14.55\%&\gray 36.40\%&\gray 21.44\%&\gray 29.60\%&\gray 19.00\%\\ \midrule
         SLMRec & 0.0343& 0.0529& 0.0226& 0.0290& 0.0429& 0.0663& 0.0288& 0.0365& 0.0292& 0.0452& 0.0196& 0.0247& 0.0309& 0.0529& 0.0165& 0.0234\\
         +SG-URInit & \textbf{0.0360$^*$}& \textbf{0.0556$^*$}& \textbf{0.0235$^*$}& \textbf{0.0307$^*$}& \textbf{0.0454$^*$}& \textbf{0.0708$^*$}& \textbf{0.0305$^*$}& \textbf{0.0388$^*$}& \textbf{0.0312$^*$}& \textbf{0.0488$^*$}& \textbf{0.0210$^*$}& \textbf{0.0265$^*$}& \textbf{0.0344$^*$}& \textbf{0.0577$^*$}& \textbf{0.0182$^*$}& \textbf{0.0256$^*$}\\
         \gray Improv. &\gray 4.96\%&\gray 5.10\%&\gray 3.98\%&\gray 5.86\%&\gray 5.83\%&\gray 6.79\%&\gray 5.90\%&\gray 6.30\%&\gray 6.85\%&\gray 7.96\%&\gray 7.14\%&\gray 7.29\%&\gray 11.33\%&\gray 9.07\%&\gray 10.30\%&\gray 9.40\%\\ \midrule
         FREEDOM & 0.0374& 0.0627& 0.0243& 0.0330& 0.0446& 0.0717& 0.0291& 0.0385& 0.0388& 0.0629& 0.0257& 0.0341& 0.0367& 0.0575& 0.0250& 0.0316\\
         +SG-URInit & \textbf{0.0391$^*$}& \textbf{0.0657$^*$}& \textbf{0.0256$^*$}& \textbf{0.0351$^*$}& \textbf{0.0471$^*$}& \textbf{0.0759$^*$}& \textbf{0.0310$^*$}& \textbf{0.0411$^*$}& \textbf{0.0406$^*$}& \textbf{0.0662$^*$}& \textbf{0.0270$^*$}& \textbf{0.0360$^*$}& \textbf{0.0404$^*$}& \textbf{0.0616$^*$}& \textbf{0.0269$^*$}& \textbf{0.0338$^*$}\\
         \gray Improv. &\gray 4.55\%&\gray 4.78\%&\gray 5.35\%&\gray 6.36\%&\gray 5.61\%&\gray 5.86\%&\gray 6.53\%&\gray 6.75\%&\gray 4.64\%&\gray 5.25\%&\gray 5.06\%&\gray 5.57\%&\gray 10.08\%&\gray 7.13\%&\gray 7.60\%&\gray 6.96\%\\ \midrule
         DRAGON & 0.0380& 0.0662& 0.0249& 0.0345& 0.0449& 0.0752& 0.0296& 0.0413& 0.0401& 0.0671& 0.0270& 0.0365& 0.0335& 0.0568& 0.0228& 0.0308\\
         +SG-URInit & \textbf{0.0396$^*$}& \textbf{0.0697$^*$}& \textbf{0.0262$^*$}& \textbf{0.0365$^*$}& \textbf{0.0473$^*$}& \textbf{0.0791$^*$}& \textbf{0.0311$^*$}& \textbf{0.0438$^*$}& \textbf{0.0426$^*$}& \textbf{0.0714$^*$}& \textbf{0.0286$^*$}& \textbf{0.0388$^*$}& \textbf{0.0379$^*$}& \textbf{0.0632$^*$}& \textbf{0.0250$^*$}& \textbf{0.0336$^*$}\\
         \gray Improv. &\gray 4.21\%&\gray 5.29\%&\gray 5.22\%&\gray 5.80\%&\gray 5.35\%&\gray 5.19\%&\gray 5.07\%&\gray 6.05\%&\gray 6.23\%&\gray 6.41\%&\gray 5.93\%&\gray 6.30\%&\gray 13.13\%&\gray 11.27\%&\gray 9.65\%&\gray 9.09\%\\ \midrule
         LGMRec & 0.0374& 0.0626& 0.0249& 0.0333& 0.0446& 0.0719& 0.0288& 0.0387& 0.0371& 0.0555& 0.0246& 0.0302& 0.0341& 0.0583& 0.0235& 0.0311\\
         +SG-URInit & \textbf{0.0394$^*$}& \textbf{0.0660$^*$}& \textbf{0.0262$^*$}& \textbf{0.0350$^*$}& \textbf{0.0472$^*$}& \textbf{0.0763$^*$}& \textbf{0.0308$^*$}& \textbf{0.0418$^*$}& \textbf{0.0399$^*$}& \textbf{0.0607$^*$}& \textbf{0.0267$^*$}& \textbf{0.0328$^*$}& \textbf{0.0376$^*$}& \textbf{0.0636$^*$}& \textbf{0.0259$^*$}& \textbf{0.0343$^*$}\\
         \gray Improv. &\gray 5.35\%&\gray 5.43\%&\gray 5.22\%&\gray 5.11\%&\gray 5.83\%&\gray 6.12\%&\gray 6.94\%&\gray 8.01\%&\gray 7.55\%&\gray 9.37\%&\gray 8.54\%&\gray 8.61\%&\gray 10.26\%&\gray 9.09\%&\gray 10.21\%&\gray 10.29\%\\ \midrule
         MENTOR & 0.0422& 0.0678& 0.0281& 0.0362& 0.0485& 0.0763& 0.0323& 0.0409& 0.0411& 0.0668& 0.0274& 0.0360& 0.0364& 0.0634& 0.0247& 0.0334\\
         +SG-URInit & \textbf{0.0444$^*$}& \textbf{0.0713$^*$}& \textbf{0.0296$^*$}& \textbf{0.0382$^*$}& \textbf{0.0511$^*$}& \textbf{0.0802$^*$}& \textbf{0.0340$^*$}& \textbf{0.0431$^*$}& \textbf{0.0432$^*$}& \textbf{0.0707$^*$}& \textbf{0.0290$^*$}& \textbf{0.0381$^*$}& \textbf{0.0463$^*$}& \textbf{0.0719$^*$}& \textbf{0.0292$^*$}& \textbf{0.0374$^*$}\\
         \gray Improv. &\gray 5.21\%&\gray 5.16\%&\gray 5.34\%&\gray 5.52\%&\gray 5.36\%&\gray 5.11\%&\gray 5.26\%&\gray 5.38\%&\gray 5.11\%&\gray 5.84\%&\gray 5.84\%&\gray 5.83\%&\gray 27.20\%&\gray 13.41\%&\gray 18.22\%&\gray 11.98\%\\
         \bottomrule
    \end{tabular}
    }
    \vskip -0.05in
\end{table*}

%% file: Tab/Dataset.tex
\begin{table}[!t]
    \centering
\caption{Statistics of all evaluation datasets.}
\small
\setlength{\tabcolsep}{1.0mm}
 \vskip -0.05in
\label{tab:dataset_statistics}
    \begin{tabular}{cccccc}
    \toprule
         \textbf{Datasets}&  \textbf{\#Users}&  \textbf{\#Items}& \textbf{\#Interactions} & \textbf{Sparsity} & \textbf{Modality}\\
         \midrule
         Baby & 19,445 & 7,050 & 160,792 & 99.88\% & V,T\\
         Sports & 35,598 & 18,357 & 296,337 & 99.95\% & V,T\\
         Clothing & 39,387 & 23,033 & 278,677 & 99.97\% & V,T\\
         TikTok & 9,319 & 6,710 & 59,541 & 99.90\% & V,T,A\\
         \bottomrule
    \end{tabular}
     \vskip -0.05in
\end{table}

%% file: Tab/Compatibility.tex
\begin{table}[!t]
\caption{Performance comparison for strategies on three datasets under Recall@5 (R@5) and NDCG@5 (N@5). +S, +A, and +G denote +SG-URInit, +AMR, and +GPT-4o, respectively.}
  \vskip -0.05in
\centering
\tabcolsep=0.04in
\label{tab:compatibility}
\resizebox{\linewidth}{!}{
    \begin{tabular}{c|c|cc|cc|cc}
    \toprule
         \multirow{2.5}{*}{Models} & Datasets& \multicolumn{2}{c|}{Baby}& \multicolumn{2}{c|}{Sports} & \multicolumn{2}{c}{Clothing}\\\cmidrule{2-8}
         & Metrics& R@5& N@5& R@5& N@5& R@5& N@5\\\midrule
         \multirow{5}{*}{MMGCN} & origin & 0.0240& 0.0160& 0.0216& 0.0143& 0.0130& 0.0110\\
         & +S & 0.0263& 0.0171& 0.0268& 0.0175& 0.0147& 0.0126\\
         & +S+A & 0.0267& 0.0174& 0.0273& 0.0177& 0.0151& 0.0128\\ 
         & +S+G & \underline{0.0271}& \textbf{0.0176}& \underline{0.0276}& \underline{0.0179}& \underline{0.0152}& \underline{0.0131}\\ 
         & +S+A+G & \textbf{0.0272}& \textbf{0.0176}& \textbf{0.0279}& \textbf{0.0180}& \textbf{0.0154}& \textbf{0.0135}\\ \midrule
         \multirow{5}{*}{LGMRec} & origin & 0.0374& 0.0249& 0.0446& 0.0288& 0.0371& 0.0246\\
         & +S & 0.0394& 0.0262& 0.0472& 0.0308& 0.0399& 0.0267\\
         & +S+A & 0.0400& 0.0267& 0.0476& 0.0311& 0.0406& 0.0272\\ 
         & +S+G & \underline{0.0404}& \underline{0.0269}& \underline{0.0481}& \underline{0.0314}& \underline{0.0410}& \underline{0.0274}\\ 
         & +S+A+G & \textbf{0.0408}& \textbf{0.0272}& \textbf{0.0484}& \textbf{0.0317}& \textbf{0.0413}& \textbf{0.0277}\\ \midrule
         \multirow{5}{*}{MENTOR} & origin & 0.0422& 0.0281& 0.0485& 0.0323& 0.0411& 0.0274\\
         & +S & 0.0444& 0.0296& 0.0511& 0.0340& 0.0432& 0.0290\\
         & +S+A & 0.0447& 0.0298& 0.0514& 0.0343& 0.0433& 0.0290\\ 
         & +S+G & \underline{0.0450}& \underline{0.0300}& \underline{0.0518}& \underline{0.0346}& \underline{0.0437}& \underline{0.0293}\\ 
         & +S+A+G & \textbf{0.0452}& \textbf{0.0303}& \textbf{0.0520}& \textbf{0.0347}& \textbf{0.0440}& \textbf{0.0296}\\ \bottomrule
    \end{tabular}
    }
    \vskip -0.05in
\end{table}

%% file: Tab/Cold_Start.tex
\begin{table}[!t]
    \centering
    \caption{Cold start analysis across Baby and Tiktok datasets.}
    \label{tab:cold-start}
    \vskip -0.05in
    \small
    \begin{tabular}{c|cc|cc}
    \toprule
        Datasets & \multicolumn{2}{c|}{Baby} & \multicolumn{2}{c}{TikTok} \\ \midrule
        Metrics & Recall@10 & NDCG@10 & Recall@10 & NDCG@10 \\ \midrule
        MMGCN & 0.0103 & 0.0062 & 0.0214 & 0.0112 \\ 
        +SG-URInit & \textbf{0.0124} & \textbf{0.0073} & \textbf{0.0250} & \textbf{0.0132} \\  \midrule
        SLMRec & 0.0172 & 0.0101 & 0.0230 & 0.0117 \\
        +SG-URInit & \textbf{0.0207} & \textbf{0.0123} & \textbf{0.0259} & \textbf{0.0133} \\  \midrule
        FREEDOM & 0.0348 & 0.0195 & 0.0301 & 0.0149 \\ 
        +SG-URInit & \textbf{0.0390} & \textbf{0.0219} & \textbf{0.0336} & \textbf{0.0166} \\ \midrule
        DRAGON & 0.0378 & 0.0212 & 0.0402 & 0.0205 \\
        +SG-URInit & \textbf{0.0419} & \textbf{0.0238} & \textbf{0.0445} & \textbf{0.0224} \\ \midrule
        LGMRec & 0.0371 & 0.0208 & 0.0333 & 0.0169 \\ 
        +SG-URInit & \textbf{0.0409} & \textbf{0.0230} & \textbf{0.0370} & \textbf{0.0187} \\ \midrule
        MENTOR & 0.0391 & 0.0219 & 0.0415 & 0.0212 \\ 
        +SG-URInit & \textbf{0.0430} & \textbf{0.0249} & \textbf{0.0459} & \textbf{0.0233} \\ \bottomrule
    \end{tabular}
    \vskip -0.05in
\end{table}

%% file: Tab/Appendix1.tex
\begin{table}[!t]
\caption{Performance comparison for weight allocation strategies on three datasets under Recall@5 (R@5) and NDCG@5 (N@5).}
 \vskip -0.05in
\centering
\tabcolsep=0.04in
\label{tab:ablation2}
\resizebox{\linewidth}{!}{
    \begin{tabular}{c|c|cc|cc|cc}
    \toprule
         \multirow{2.5}{*}{Models} & Datasets& \multicolumn{2}{c|}{Baby}& \multicolumn{2}{c|}{Sports} & \multicolumn{2}{c}{Clothing}\\\cmidrule{2-8}
         & Metrics& R@5& N@5& R@5& N@5& R@5& N@5\\\midrule
         \multirow{4}{*}{MMGCN} & origin & 0.0240& 0.0160& 0.0216& 0.0143& 0.0130& 0.0110\\
         & SG-URInit & \textbf{0.0263}& \textbf{0.0171}& 0.0268& 0.0175& \textbf{0.0147}& \textbf{0.0126}\\
         & -bi-degree & 0.0261& 0.0170& \textbf{0.0269}& \textbf{0.0177}& 0.0141& 0.0123\\
         & -equal & 0.0258& 0.0168& 0.0261& 0.0170& 0.0140& 0.0122\\
          \midrule
         \multirow{4}{*}{LGMRec} & origin & 0.0374& 0.0249& 0.0446& 0.0288& 0.0371& 0.0246\\
         & SG-URInit & \textbf{0.0394}& \textbf{0.0262}& \textbf{0.0472}& \textbf{0.0308}& \textbf{0.0399}& \textbf{0.0267}\\
         & -bi-degree & 0.0391& 0.0260& 0.0468& 0.0305& 0.0395& 0.0264\\
         & -equal & 0.0388& 0.0258& 0.0466& 0.0303& 0.0391& 0.0262\\\midrule
         \multirow{4}{*}{MENTOR} & origin & 0.0422& 0.0281& 0.0485& 0.0323& 0.0411& 0.0274\\
         & SG-URInit & \textbf{0.0444}& \textbf{0.0296}& \textbf{0.0511}& \textbf{0.0340}& \textbf{0.0432}& \textbf{0.0290}\\
         & -bi-degree & 0.0442& 0.0295& 0.0507& 0.0337& 0.0427& 0.0286\\
         & -equal & 0.0438& 0.0292& 0.0503& 0.0334& 0.0424& 0.0284\\\bottomrule
    \end{tabular}
    }
    \vskip -0.05in
\end{table}

%% file: Tab/Cost.tex
\begin{table}[!t]
    \centering
\caption{Time cost for user initialization.}
 \vskip -0.05in
\label{tab:cost}
    \begin{tabular}{ccccc}
    \toprule
         Datasets& Baby& Sports& Clothing& TikTok\\
         \midrule
         $K = 4$& 15.23s& 37.02s& 43.01s& 4.75s\\ 
         $K = 8$& 17.88s& 42.30s& 49.32s& 5.40s\\ 
         \bottomrule
    \end{tabular}
     \vskip -0.05in
\end{table}

%% file: Tab/Appendix2.tex
\begin{table*}[!th]
\caption{Analysis for dynamic $K$ for different modalities on two datasets under Recall@5 (R@5) and NDCG@5 (N@5).}
 \vskip -0.05in
\centering
\tabcolsep=0.04in
\label{tab:ablation3}
    \begin{tabular}{c|c|ccccc|ccccc|ccccc|ccccc}
    \toprule
         \multirow{2.5}{*}{Models} & Variant & \multicolumn{5}{c|}{Baby} & \multicolumn{5}{c|}{Sports}& \multicolumn{5}{c|}{Clothing}& \multicolumn{5}{c}{TikTok}\\\cmidrule{2-22}
         & Metrics& R@5& N@5& T& V& A & R@5& N@5& T& V& A & R@5& N@5& T& V& A & R@5& N@5& T& V& A \\\midrule
         \multirow{3}{*}{MMGCN} & Origin & 0.0240& 0.0160& -& -& -& 0.0216& 0.0143& -& -& -& 0.0130& 0.0088& -& -& -& 0.0250& 0.0125& -& -& -\\
         & SG-URInit & 0.0263& 0.0171& 4& 4& -& 0.0268& 0.0175& 4& 4& -& \textbf{0.0147}& \textbf{0.0100}& 8& 8& -& 0.0341& 0.0162& 4& 4& 4\\
         & Dynamic $K$ & \textbf{0.0270}& \textbf{0.0174}& 8& 4& -& \textbf{0.0271}& \textbf{0.0177}& 8& 4& -& \textbf{0.0147}& \textbf{0.0100}& 8& 8& -& \textbf{0.0352}& \textbf{0.0168}& 8& 4& 8\\ \midrule
         \multirow{3}{*}{FREEDOM} & Origin & 0.0374& 0.0243& -& -& -& 0.0446& 0.0291& -& -& -& 0.0388& 0.0257& -& -& -& 0.0367& 0.0250& -& -& -\\
         & SG-URInit & 0.0391& 0.0256& 4& 4& -& 0.0471& 0.0310& 4& 4& -& \textbf{0.0406}& \textbf{0.0270}& 8& 8& -& 0.0404& 0.0269& 4& 4& 4\\
         & Dynamic $K$ & \textbf{0.0395}& \textbf{0.0259}& 8& 4& -& \textbf{0.0477}& \textbf{0.0314}& 8& 4& -& \textbf{0.0406}& \textbf{0.0270}& 8& 8& -& \textbf{0.0412}& \textbf{0.0274}& 8& 4& 8\\ \midrule
         \multirow{3}{*}{DRAGON} & Origin & 0.0380& 0.0249& -& -& -& 0.0449& 0.0296& -& -& -& 0.0401& 0.0270& -& -& -& 0.0335& 0.0228& -& -& -\\
         & SG-URInit & 0.0396& 0.0262& 4& 4& -& 0.0473& 0.0311& 4& 4& -& \textbf{0.0426}& \textbf{0.0286}& 8& 8& -& 0.0379& 0.0250& 4& 4& 4\\
         & Dynamic $K$ & \textbf{0.0402}& \textbf{0.0266}& 8& 4& -& \textbf{0.0480}& \textbf{0.0314}& 8& 4& -& \textbf{0.0426}& \textbf{0.0286}& 8& 8& -& \textbf{0.0390}& \textbf{0.0256}& 8& 4& 8\\ \midrule
         \multirow{3}{*}{MENTOR} & Origin & 0.0422& 0.0281& -& -& -& 0.0485& 0.0323& -& -& -& 0.0411& 0.0274& -& -& -& 0.0364& 0.0247& -& -& -\\
         & SG-URInit & 0.0444& 0.0296& 4& 4& -& 0.0511& 0.0340& 4& 4& -& \textbf{0.0432}& \textbf{0.0290}& 8& 8& -& 0.0463& 0.0292& 4& 4& 4\\
         & Dynamic $K$ & \textbf{0.0452}& \textbf{0.0301}& 8& 4& -& \textbf{0.0515}& \textbf{0.0344}& 8& 4& -& \textbf{0.0432}& \textbf{0.0290}& 8& 8& -& \textbf{0.0471}& \textbf{0.0296}& 8& 4& 8\\ \bottomrule
    \end{tabular}
    \vskip -0.05in
\end{table*}

%% file: Tex/5.Related_Work.tex
\section{Related Work}
Many recent studies have incorporated multimodal information to address the data sparsity problem in recommendation systems. Early work such as VBPR \cite{he2016vbpr} leveraged visual content alongside matrix factorization, while subsequent studies \cite{chen2019personalized,liu2019user,yu2023multi,xu2025best,xu2025enhancing,xu2025vi} integrated both visual and textual modalities for improved item representation. Evolving from traditional architectures, methods like MMGCN \cite{wei2019mmgcn} and GRCN \cite{wei2020graph} employed graph convolutional networks (GCNs) to model user-item interactions, with GRCN refining edge pruning to reduce noise. Other approaches such as DualGNN \cite{wang2021dualgnn}, LATTICE \cite{zhang2021mining}, and FREEDOM \cite{zhou2023tale} enhanced user and item representations through co-occurrence graphs, semantic graphs, and stabilized item representations, respectively. Further advancements include contrastive learning-based models like MMSSL \cite{wei2023multi} and MICRO \cite{zhang2022latent}, modal fusion techniques in BM3 \cite{zhou2023bootstrap} and MENTOR \cite{xu2025mentor}, and innovations using hypergraphs and diffusion models in LGMRec \cite{guo2024lgmrec} and DiffMM \cite{jiang2024diffmm}. COHESION \cite{xu2025cohesion} further introduced dual-stage modality fusion to enhance user and item representations.

Despite these advancements, existing works largely overlook the critical role of user representation initialization. Unlike items, which are enriched with modality information, users lack such inherent features. This disparity results in a semantic gap between meaningful item representations and randomly initialized user representations. To this end, we propose Semantically Guaranteed User Representation Initialization (SG-URInit). SG-URInit integrates modality features of users' interacted items and cluster-level global features to construct semantically guaranteed user representations. SG-URInit effectively captures both local and global semantics, providing a robust initialization for multimodal recommendations.

%% file: Tex/6.Conclusion.tex
\section{Conclusion}
In this paper, we aim to address the critical semantic gap arising from random user initialization in multimodal recommendation, which further introduces noise during graph convolution. We propose SG-URInit, a training-free and model-agnostic semantically guaranteed user representation initialization that integrates interacted items' modality information and cluster-level information, effectively capturing both local (item-level) and global (cluster-level) semantics. Extensive experiments on multiple real-world datasets across various advanced models demonstrate that SG-URInit significantly boosts recommendation performance, alleviates the item cold-start problem, and accelerates convergence, offering an efficient and practical solution for multimodal recommendations.